\documentclass[jgr]{agutex}

 \usepackage{lineno}
  \usepackage{graphicx}
  \setkeys{Gin}{draft=false}
\usepackage{color}

\authorrunninghead{EDBERG ET AL.}

\titlerunninghead{CIR IMPACTS ON COMET 67P}

\authoraddr{N. J. T. Edberg, A. I. Eriksson, E. Odelstad, E. Vigren, D. J. Andrews, F. Johansson, Swedish Institute of Space Physics, Uppsala, Box 537, SE-75121, Sweden}
\authoraddr{J. L. Burch, R. Goldstein, K. Mandt, P. Mokashi, Southwest Research Institute, 6220 Culebra Rd., San Antonio, TX 78238, USA}
\authoraddr{C. Carr, E. Cupido, Imperial College London, Exhibition Road, London SW7 2AZ, United Kingdom}
\authoraddr{K.-H. Glassmeier, C. Koenders, I. Richter, TU - Braunschweig, Institute for Geophysics and extraterrestrial Physics, Mendelssohnstr. 3} 
\authoraddr{J. Halekas, University of Iowa, Iowa City, Iowa, USA}
\authoraddr{P. Henri, Laboratoire de Physique et Chimie de l'Environnement et de l'Espace, OrlŽans Cedex 2, France}
\authoraddr{Z. Nemeth, Wigner Research Institute, Budapest, HUngary}
\authoraddr{H. Nilsson, G. Stenberg Wieser, R. Ramstad, Swedish Institute of Space Physics, Box 812, 981 28 Kiruna, Sweden}

\begin{document}

%
%
\title{Solar wind interaction with comet 67P: impacts of corotating interaction regions}

%
%
\authors{N. J. T. Edberg, \altaffilmark{1}
A. I. Eriksson, \altaffilmark{1}
E. Odelstad, \altaffilmark{1,2}
E. Vigren, \altaffilmark{1}
D. J. Andrews, \altaffilmark{1}
F. Johansson, \altaffilmark{1}
J. L. Burch, \altaffilmark{3}
C. M. Carr, \altaffilmark{4}
E. Cupido, \altaffilmark{4}
K.-H. Glassmeier, \altaffilmark{5}
R. Goldstein,\altaffilmark{3}
J. S. Halekas,\altaffilmark{6}
P. Henri, \altaffilmark{7}
C. Koenders, \altaffilmark{5}
K. Mandt, \altaffilmark{3}
P. Mokashi, \altaffilmark{3} 
Z. Nemeth, \altaffilmark{8} 
H. Nilsson, \altaffilmark{9}
R. Ramstad, \altaffilmark{9}
I. Richter, \altaffilmark{5}
G. Stenberg Wieser \altaffilmark{9}
}

\altaffiltext{1}{Swedish Institute of Space Physics, Uppsala, Sweden (ne@irfu.se)}
\altaffiltext{2}{Department of Physics and Astronomy, Uppsala University, Uppsala, Sweden}
\altaffiltext{3}{Southwest Research Institute, San Antonio, USA}
\altaffiltext{4}{Space and Atmospheric Physics Group, Imperial College London, UK} 
\altaffiltext{5}{TU - Braunschweig, Institute for Geophysics and Extraterrestrial Physics, Braunschweig, Germany}
\altaffiltext{6}{Department of Physics and Astronomy, University of Iowa, Iowa City, IA, USA}
\altaffiltext{7}{Laboratoire de Physique et Chimie de l'Environnement et de l'Espace, Orleans, France}
\altaffiltext{8}{Wigner Research Center for Physics, Budapest, Hungary}
\altaffiltext{9}{Swedish Institute of Space Physics, Kiruna, Sweden}









%
%


\begin{abstract}
We present observations from the Rosetta Plasma Consortium of the effects of stormy solar wind on comet 67P/Churyumov-Gerasimenko. Four corotating interaction regions (CIRs), where the first event has possibly merged with a CME, are traced from Earth via Mars (using Mars Express and MAVEN) and to comet 67P from October to December 2014. When the comet is 3.1-2.7 AU from the Sun and the neutral outgassing rate $\sim10^{25}-10^{26}$ s$^{-1}$ the CIRs significantly influence the cometary plasma environment at altitudes down to 10-30 km. The ionospheric low-energy \textcolor{black}{($\sim$5 eV) plasma density increases significantly in all events, by a factor $>2$ in events 1-2 but less in events 3-4. The spacecraft potential drops below -20V upon impact when the flux of electrons increases}. The increased density is \textcolor{black}{likely} caused by compression of the plasma environment, increased particle impact ionisation, and possibly charge exchange processes and acceleration of mass loaded plasma back to the comet ionosphere. During all events, the fluxes of suprathermal ($\sim$10-100 eV) electrons increase significantly, suggesting that the heating mechanism of these electrons is coupled to the solar wind energy input. At impact the magnetic field strength in the coma increases by a factor of ~2-5 as more interplanetary magnetic field piles up around of the comet. During two CIR impact events, we observe possible plasma boundaries forming, or moving past Rosetta, as the strong solar wind compresses the cometary plasma environment. \textcolor{black}{We also discuss the possibility of seeing some signatures of the ionospheric response to tail disconnection events}.
\end{abstract}

%
%

%

\begin{article}

%
%

\section{Introduction}
The interaction of the solar wind with a cometary plasma environment has previously only been studied in situ during satellite flybys. With Rosetta in orbit around comet 67P/Churyumov-Gerasimenko (hereafter 67P), it is now possible to study the interaction over longer periods of time, and especially to study how the cometary magnetosphere responds to variations in the solar wind. 

Rosetta arrived at comet 67P on 6 August 2014, to begin conducting continuous measurements of the near-nucleus ($<100$ km) plasma environment as well as of the neutral gas and dust. During the autumn of 2014 Rosetta gradually and slowly approached the comet, moving with a velocity on the order of 1 m/s, and reaching a minimum altitude of about 10 km during October. The comet lowered its heliocentric distance from 3.1 to 2.7 AU, in the interval from 1 Oct 2014 - 1 Jan 2015.

At the arrival of Rosetta in August 2014, the comet was relatively inactive and had just begun to form an \textcolor{black}{induced magnetosphere in the solar wind \citep{nilsson2015}. The neutral outgassing rate from the comet was measured to be $\sim10^{25}$ s$^{-1}$ \citep{gulkis2015}. As the emitted neutral gas becomes ionised, predominantly through photoionization, particle impacts and/or charge exchange processes \citep{cravens1987, burch2015, vigren2015}  an ionosphere, which initially expands radially outward, forms around the comet nucleus. As the comet approaches the sun, the ionosphere grows and becomes denser, and the induced magnetosphere concurrently expands.} The density of the ionospheric plasma close to the nucleus decreases with distance as $1/R$ \citep{edberg2015}, in agreement with theory simplified by the neglect of field influence and chemical loss \citep{vigren2015}. 
 
Pickup ions of cometary origin have been observed since arrival and have evolved with time to become more energetic closer to the Sun \citep{goldstein2015,nilsson2015b}. The outgassing of neutrals was found to be quite inhomogeneous and most of the gas was observed over the neck region of the comet \citep{hassig2015}. \citet{edberg2015} mapped the ionospheric plasma and found it to be distributed similarly as the neutrals, indicating that plasma from local ionisation of neutrals (in the region between the nucleus and Rosetta) dominated the cold plasma environment. This also implies that the ionospheric structure is modulated by the comet spin period. Furthermore, 67P was found to have a plasma environment that is full of small scale density and magnetic field variations and where instabilities and waves are common \citep{richter2015}, complicating the interaction with the solar wind. \textcolor{black}{\citet{volwerk2016} reported on observations of mirror-mode waves that were generated following a compression of the plasma environment due to increased solar wind dynamic pressure}. A population of suprathermal electrons ($\sim10-100$ eV) has been observed continuously after arrival, although with significant variations in both energy and flux over time \citep{clark2015}. 

Comet 67P is a relatively weakly outgassing comet and no \textcolor{black}{bow shock, diamagnetic cavity or ionopause, for instance,} was observed in the interval covered in this paper, i.e. until 1 Jan 2015. Some piling-up of the interplanetary magnetic field (IMF) was occurring but no constant pile-up boundary was observed, which is in agreement with predictions from simulations \citep{koenders2013, rubin2014}. The ion instruments on Rosetta were able to measure the solar wind more or less continuously since arrival (until the end of March of 2015 when the coma had grown), although significant deflection of the solar wind, by more than 45 degrees, was observed \citep{nilsson2015b, broiles2015}. 

In this paper we will specifically study the response of the cometary plasma environment to stormy space weather, when impacts of corotating interaction regions (CIR) occur. The purpose is to identify the main effects these solar wind pressures pulses have on the comet ionosphere. 

CIRs form when slow solar wind is caught up by faster solar wind and an interaction region forms in between the two flows, typically characterised by an increased magnetic field strength, plasma density and pressure. Across the CIR, the magnetic field polarity often reverses. The slow and fast solar winds emanate from specific regions on the Sun, which can be emitting plasma similarly during several solar rotations (one solar rotation being approximately 27 days) such that an interaction region is continuously formed. As the Sun rotates, this interaction region will sweep across the heliosphere while also expanding radially outward, until it eventually impacts on any planet, moon or comet in its way, where it most likely will cause a significant disturbance to the plasma environment. At Mars and Venus, CIR impacts have been shown to increase the ionospheric escape rates\textcolor{black}{ \citep[e.g][]{dubinin2008,edberg2009b,edberg2010,edberg2011b}}. CIRs are typically less impulsive than coronal mass ejections (CME), but are instead a more common and regular phenomena in interplanetary space since they emanate from coronal holes that are more stable in time. CMEs propagate mainly radially outward (with some angular spread), but can often merge with any slower CIR ahead of it. 

CME impacts on comets have been observed to lead to so called tail disconnection events, when oppositely directed magnetic fields in the tail of the comet reconnect with each other \textcolor{black}{\citep{vourlidas2007}}. A plasmoid then forms which disconnects from the comet \citep{niedner1978}. Similar events could also occur during CIR impact events. Since Rosetta is always orbiting relatively close to the nucleus any tail disconnection event happening far down the tail will, most likely, not be directly observed by Rosetta, but the effects of it closer to the nucleus might \textcolor{black}{still} be observed. 

We will present measurements from the plasma environment around comet 67P during 4 separate CIR impacts to study how the comet magnetosphere is affected during such events. We have limited our study to 3 months (1 October 2014 - 1 January 2015) when the comet was at heliocentric distances of 3.1-2.7 AU and relatively inactive (outgassing $\sim10^{25}-10^{26}$ s$^{-1}$). The study is naturally limited in space \textcolor{black}{and time to where Rosetta was located in this interval, i.e. in orbit at an altitude of 10-30 km and close to the terminator plane}. The paper is organised as follows: first we introduce the instruments used in this paper, followed by a section describing CIR \textcolor{black}{propagation from} Earth, via Mars and finally to 67P. We then present measurements from Rosetta at 67P during the CIR impacts, and finish with a discussion and summary.

\section{Instruments}
The Rosetta spacecraft carries five instruments for measuring plasma properties as well as magnetic and electric fields around the comet. These form the Rosetta Plasma Consortium (RPC) \citep{carr2007}. In this paper we will present a combined data set from all instruments. Each individual instrument is briefly described below.

The Langmuir probe instrument (LAP) \citep{eriksson2007} consists of two spherical Langmuir probes (LAP1 and LAP2), mounted on booms, 2.2 m and 1.6 m long, respectively, from hinge to probe. Here we will mainly use data from LAP1 when in `sweep' mode, to obtain the ion and electron density as well as the spacecraft potential at a cadence of normally 96 s or 160 s.  From the Langmuir probe sweeps both the electron and ion density can be obtained. However, when the spacecraft potential is very negative (which it often is during the interval studied in this paper \textcolor{black}{due to high fluxes of $\sim$5 eV electrons}) the electrons are to a large extent accelerated away from the probe and the electron density obtained by the probe sweeps is underestimated. Regarding the ions, the instrument sweep range has not always been set to sufficiently low negative values to fully sample the undisturbed ion current and the ion density estimate is then more uncertain. During intervals of very negative spacecraft potential the ions are fortunately more easily attracted by the probe. Still, this sweep range limitation leads us to use the sweep derived electron density estimate during intervals when the sweep is limited to above -5 V, while otherwise we use the ion density estimate. Furthermore, we are also able to use the measured spacecraft potential obtained from the sweeps as a proxy for the electron density. This density estimate is also dependent on the electron temperature, which most of the time is about 5 eV \textcolor{black}{\citep{odelstad2015}}. In addition to the sweep mode, LAP can also be run with a fixed bias-potential, where the probe attracts an ion (or electron) current when the probe potential is set to a positive (negative) voltage. When possible, the LAP-derived densities are crosschecked with data from the mutual impedance probe (MIP), described next. 

The MIP instrument \citep{trotignon2007} consists of two receiving and two transmitting electrodes, mounted on the same boom as LAP1. MIP is able to retrieve the electron density from the position of the electron plasma frequency in the mutual impedance spectra. MIP can operate in two different ways: in short Debye length (SDL) mode by using one or two of its own transmitters situated at 40 and 60 cm from the receivers, or in long Debye length (LDL) mode when using the LAP2 probe as transmitter, at a distance of about 4 m from the MIP receivers. The SDL and LDL modes cannot be run simultaneously.

On the one hand, when the spacecraft potential becomes more negative than the LAP sweep range, it is not possible to obtain the electron density from the LAP sweeps. On the other hand, MIP can retrieve plasma parameters only when (i) the ratio of the transmitter-receiver baseline length to the Debye length is large enough, and (ii) when the electron plasma frequency is in the operated MIP frequency range, which is limited to 7-168 kHz in the LDL mode. This explains why, in the early stage of the low comet activity with little cooling of electrons, there is a gap (about 350 - 1000 cm$^{-3}$) between the density ranges covered by the LDL and SDL  modes. Comparing the observations from the MIP and LAP experiments allows us to be more confident about the plasma density estimates.

The magnetometer (MAG) \citep{glassmeier2007b} uses two triaxial fluxgate sensors mounted on the  same boom as LAP2, and provides vector measurements of the magnetic field. Here we use 1 min averages from the outboard sensor,\textcolor{black}{which is located 15 cm further out on the boom (and therefore} less affected by the spacecraft generated fields). The magnetic field is shown in the comet centered solar equatorial coordinate system (CSEQ). In this system the x-axis points from the comet to the Sun, the z-axis is the component of the Sun's north pole orthogonal to the x-axis, and the y-axis completes the right-handed reference frame.

The ion and electron sensor (IES) \citep{burch2007} consists of  two electrostatic plasma analyzers, one for ions and one for electrons. Both analyzers measure in the energy/charge range 1 eV/q - 18 keV/q in 128 steps with a resolution of 8$\%$ and a field of view of $90^\circ \times 360^\circ$. The angular resolution is $5^\circ \times 22.5^\circ$ for electrons and $5^\circ \times 45^\circ$ for ions. Of particular interest here, the anodes facing the solar wind flow are further segmented into $5^\circ \times 5^\circ$ sectors. A full 3D scan of the instrument typically takes 256 s.

The ion composition analyzer (ICA) \citep{nilsson2007} also measures the ion distribution, in the energy range 10 eV - 40 keV with a field of view of $90^\circ \times 360^\circ$, and can in addition resolve masses with a resolution high enough to separate e.g. protons, helium and water group ions. The angular resolution is $5^\circ \times 22.5^\circ$ and a full 3D scan takes 192 s. For technical reasons, ICA was only on intermittently during the first months after arrival and began more continuous operations in 2015. This unfortunately leaves many data gaps for the events presented in this paper.

Furthermore, in order to monitor the solar wind density and velocity for tracking solar wind structures propagating toward Rosetta and comet 67P, we make use of solar wind measurements from the ACE spacecraft, at Earth's first Lagrange point, as well as measurements from the particle instruments on Mars Express and the Mars Atmosphere and Volatile Evolution mission (MAVEN) in orbit around Mars. Mars Express and MAVEN both spend parts of their orbits in the undisturbed solar wind and can then measure its density and velocity \citep{halekas2013,ramstad2015}. Mars Express carries the analyzer of space plasmas and energetic ions (ASPERA-3) \citep{barabash2006}, while MAVEN carries the solar wind ion analyzer (SWIA) \citep{halekas2013}.

\section{Observations}
\subsection{CIR observations at Earth, Mars and 67P}
\begin{figure}
 \noindent\includegraphics[width=12cm]{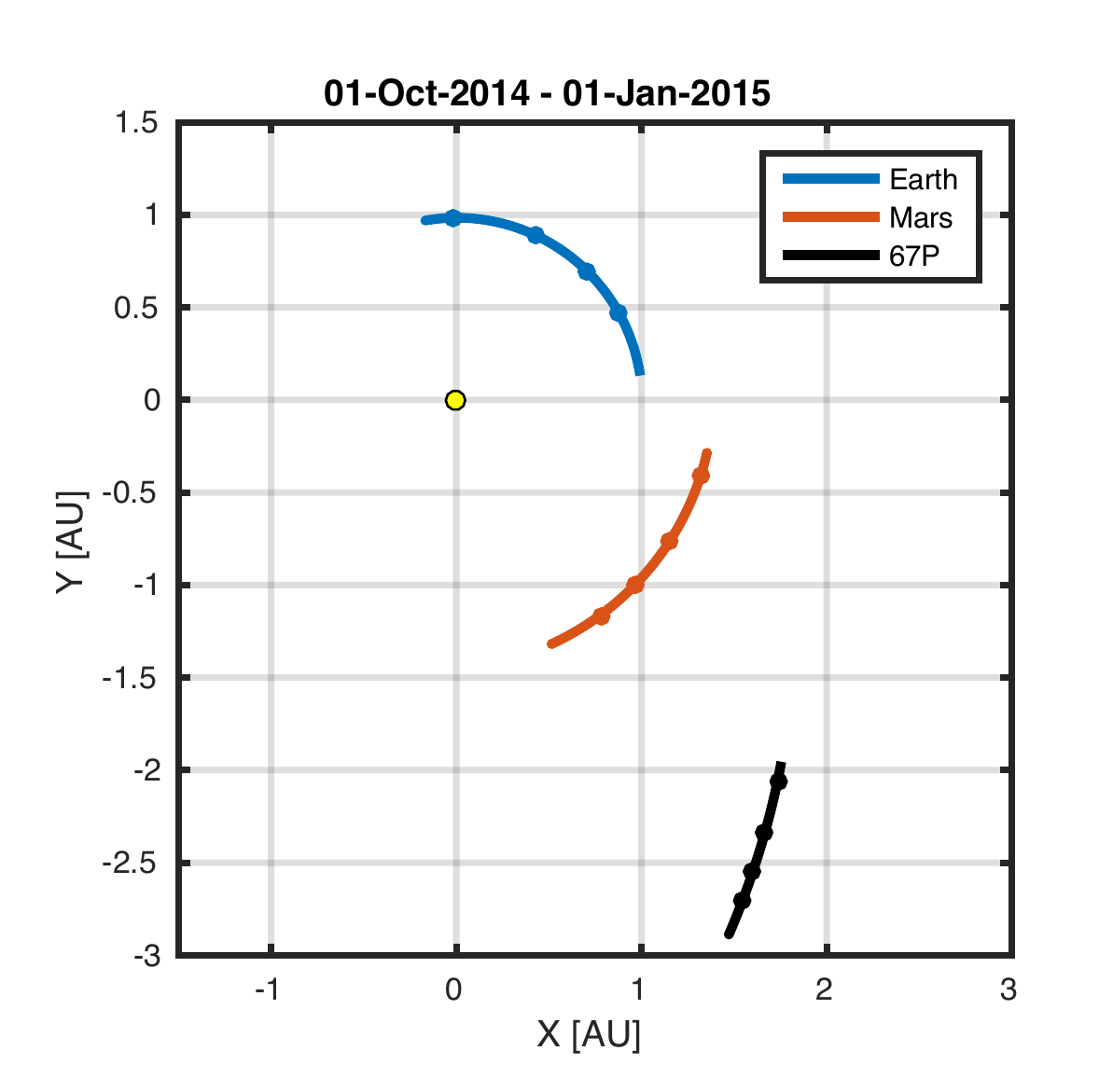}
 \caption{Positions of Earth, Mars and comet 67P during three months in late 2014 in ecliptic J2000 coordinates. \textcolor{black}{In this interval the passings of 4 CIRs were observed from solar wind measurements at each celestial body. The times of impacts are indicated by filled circles}.}
 \label{fig:position}
 \end{figure}

Figure \ref{fig:position} shows the positions of Earth, Mars and comet 67P from 1 October 2014 to 1 January 2015. Earth was ahead of 67P in heliospheric longitude by more than 45$^\circ$ during the entire interval, while Mars was more or less in the same longitude sector, and in that respect quite suitable for monitoring the solar wind upstream of the comet. However, since the radial distance between Mars and comet 67P was at least 1.2 AU during the interval studied in this paper, any solar wind structure passing by Mars might evolve significantly in both density and velocity profile before reaching 67P.  

\begin{figure}
  \noindent\includegraphics[width=19cm]{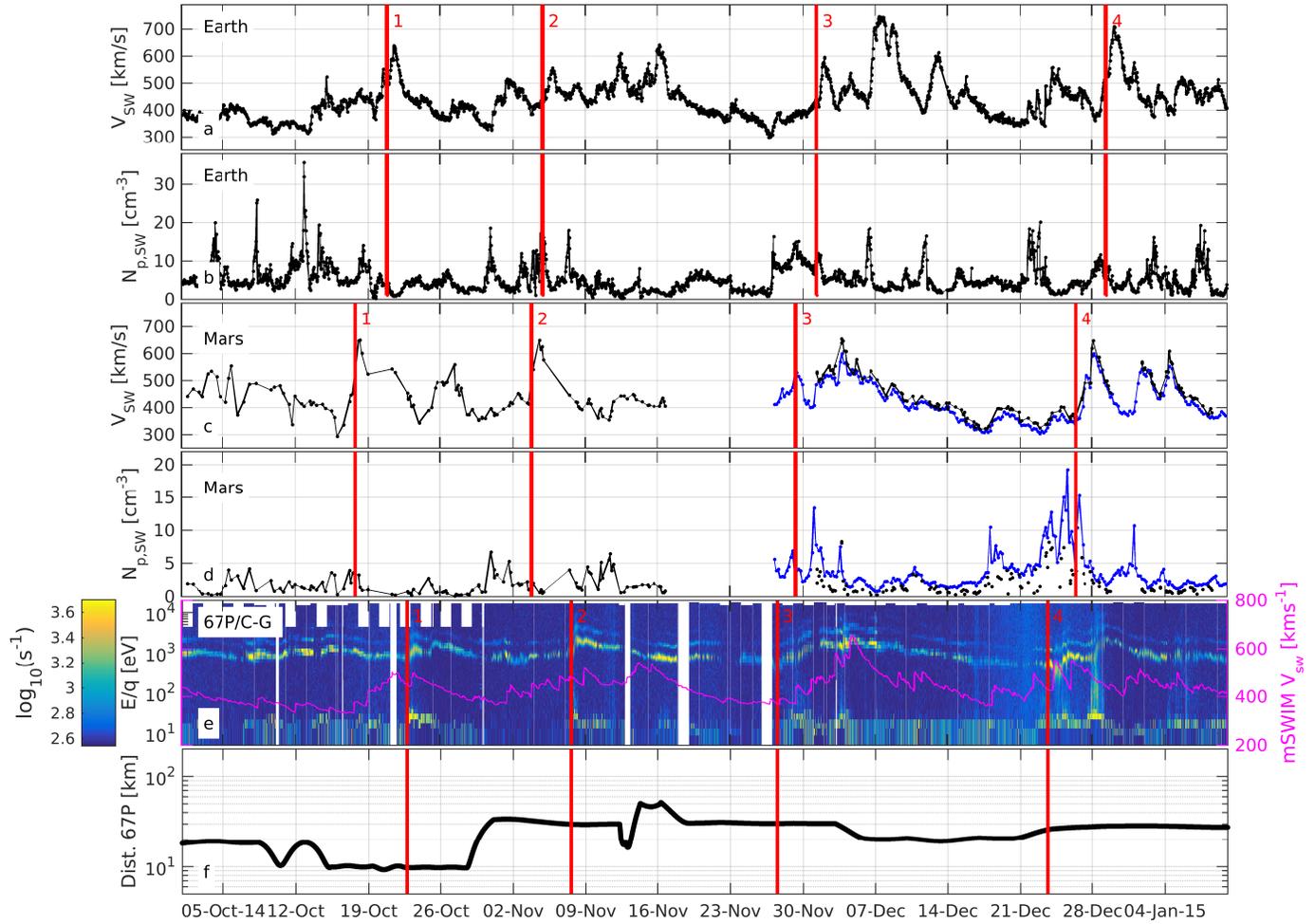}
 \caption{Time series of (a and b) solar wind density and speed from ACE at Earth, (c and d) solar wind speed and density from Mars Express (black) and MAVEN (blue) at Mars, (e) \textcolor{black}{Rosetta/IES ion spectrogram (summed over all azimuths and sectors), together with the  solar wind velocity from the mSWIM model (magenta)} and (f) distance between Rosetta and the comet centre of mass. The red vertical bars indicate the impact time of the CIRs, as determined from \textcolor{black}{the Rosetta data}.}
 \label{fig:sw}
 \end{figure}

Between 1 October 2014 and 1 January 2015 four events of solar wind pressure pulses (high solar wind velocity, plasma fluxes, pressure, magnetic field) were observed in satellite data from each body. These four events are shown in Figure \ref{fig:sw} and are indicated by red vertical lines in ACE data (Figure \ref{fig:sw}a-b), in Mars Express and MAVEN data (Figure \ref{fig:sw}c-d) and in Rosetta data (Figure \ref{fig:sw}e). \textcolor{black}{The start of each event is identified by eye from the data, primarily from the when the solar wind velocity starts to increase}. At 67P, the solar wind proton energy (which can be translated to velocity) is indicated by the bright yellow line at about 1 keV, in Figure \ref{fig:sw}e. Note that the solar wind density is not easily derived from Rosetta data as the S/C environment is dominated by plasma of cometary origin \citep{edberg2015}. \textcolor{black}{In Figure \ref{fig:sw}e we also show modeled solar wind velocity from the Michigan Solar Wind Model (mSWIM) model \citep{zieger2008}. This model uses an 1.5-D MHD code to propagate solar wind parameters from Earth out to Rosetta. Although the alignment between the Earth and 67P is not ideal for this comparison, there is a general good agreement between the model results and the IES data on how the solar wind velocity varies in this interval}. 

The CIR structures impact on comet 67P on 22 Oct, 7 Nov, 27 Nov and 22 Dec 2014. Each event can be fairly easily traced back toward the Sun using the Mars Express/MAVEN and ACE solar wind monitoring data. The fact that it is possible to track them to Earth and Mars despite the large longitudinal separation ($\sim$70$^\circ$ between Earth and 67P), indicates that these structures (at least the last three events) are CIRs. The first event is in fact probably a CIR that has merged with a coronal mass ejection (CME), observed to have been ejected from the Sun on 14 Oct 2014. In any case, in terms of velocity and density enhancement there is not much difference during the merged CME/CIR event and the three CIRs. For simplicity, we will therefore refer to the four cases as CIRs.

The four CIRs pass by Mars and Earth approximately when expected, as determined from when they are observed to pass by 67P. If assuming that these structures are CIRs, they should propagate radially outward with the solar wind speed and sweep interplanetary space with the Sun's rotation speed. To estimate the delay time from one location in interplanetary space to another, one has to take into account both the radial and the longitudinal time delay. 

The radial propagation time from Earth to 67P at 3.1 AU is about 9 days, and to 67P when at 2.7 AU about 7.3 days, if assuming a radial solar wind velocity of 400 kms$^{-1}$. The longitudinal time difference between Earth and 67P in early October (the time it takes the Sun to turn 70$^\circ$) is about 5.3 days assuming a solar rotation period of 27 days. Any solar wind structure passing by Earth in October should arrive at the comet roughly $9-5.3 = 3.7$ days later. Similarly, a solar wind structure passing by Earth in late December (when Earth and 67P are separated by almost 150$^\circ$, i.e. 11.2 days) should have arrived at the comet about $7.3-11.2 = 3.9$ days earlier. This matches up \textcolor{black}{reasonably} well with the velocity structures indicated by the red lines in the time series of solar wind data in Figure \ref{fig:sw}a and Figure \ref{fig:sw}e. \textcolor{black}{The actual time delay between the events observed at Earth and at 67P are 3.1 days, 1.1 days, 2.0 days and 3.0 days, respectively.}

If comparing Mars and 67P, a solar wind structure travelling at 400 kms$^{-1}$ that passes by Mars in October should arrive roughly 7 days later at 67P, when almost radially aligned. In late December, when the longitudinal separation between Mars and 67P has increased to about 40$^\circ$, a CIR would be seen about 3 days earlier at 67P. This also seems to be the case from the comparison of arrival times of velocity peaks in Figure \ref{fig:sw}c and Figure \ref{fig:sw}e.  Finally, comparing ACE and Mars one notes that during the entire autumn there is a $\sim90^\circ$ longitudinal separation between Earth and Mars, meaning that a CIR would be observed first at Mars and then about 4 days later at ACE.
 
The individual CIRs are not exactly separated in time by one solar rotation, suggesting that the events are really different CIRs, which originate from different locations on the Sun or that the source region on the Sun has changed. Also, when reaching the heliocentric distance of 67P the original events seen at ACE have possibly had time to evolve and merge with other solar wind structures. If looking again at the time series of ACE data, Figure \ref{fig:sw}a, 
we note e.g. that the second CIR observed in the ACE data, on 5 November, is surrounded by three additional velocity peaks with successively increasing velocity from 1 - 16 November. These four peaks only show up as one single structure at Mars on 4 November, indicating that these structures have either merged, the source region on the Sun has changed or, possibly, that some of the smaller peaks seen by ACE were not CIRs but rather smaller CMEs. 

In summary, tracking CIRs from Earth out to 67P at 3 AU and predicting their arrival times is possible, although somewhat uncertain and their definition to some extent subjective. Nevertheless, and most importantly for this study, in Rosetta RPC-IES data four clear events are observed in the ion spectrogram and manifested as sudden increases in the solar wind flux and energy, which are interpreted as CIR impacts.

\begin{figure}
 \noindent\includegraphics[width=14cm]{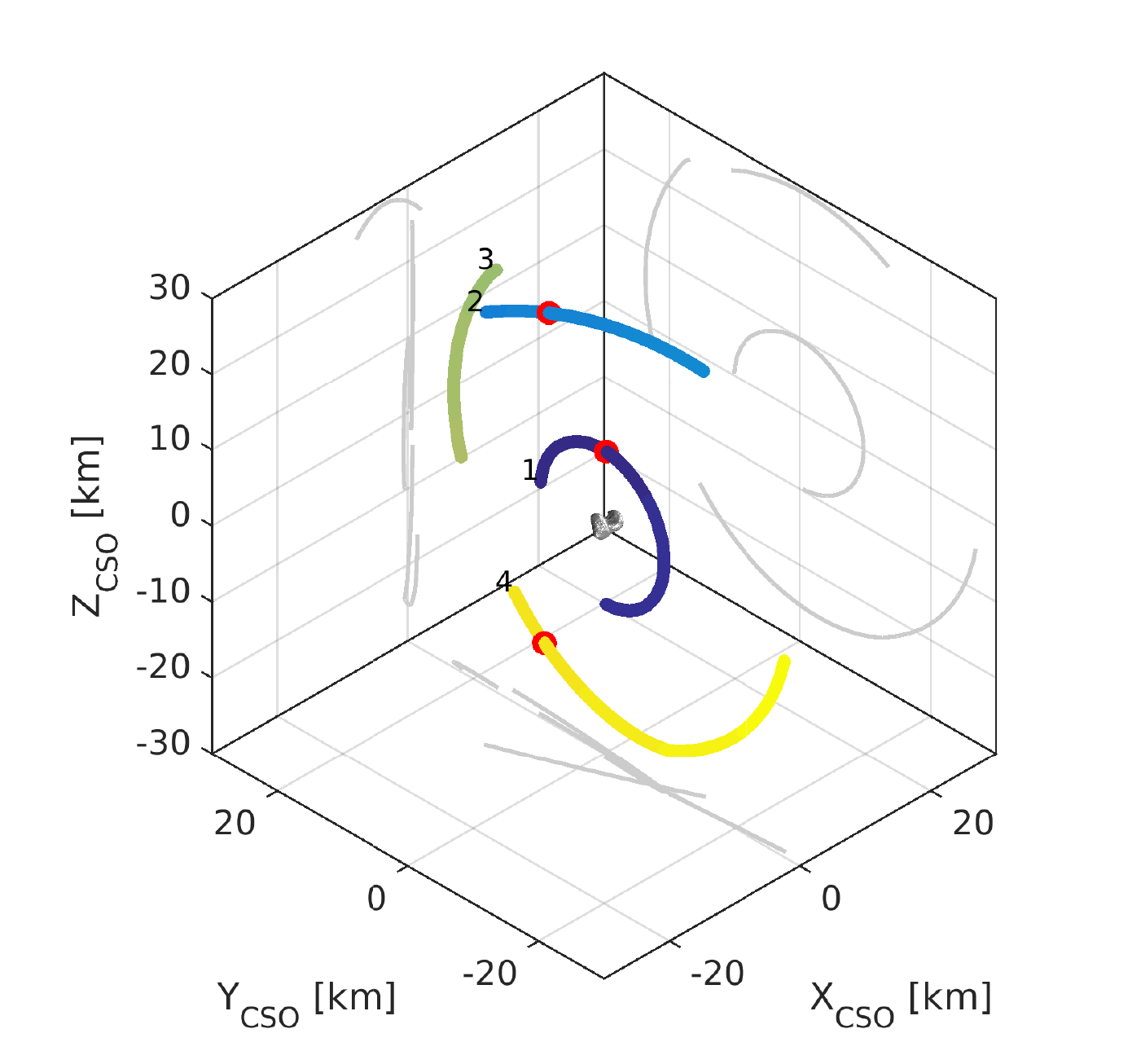}
 \caption{The position of Rosetta in the comet-centred CSO reference frame during the four CIR impacts. \textcolor{black}{The sun is toward positive $X_{CSO}$. }The positions are shown for the same intervals as the time series in Figures \ref{fig:eventa}-\ref{fig:eventd}.The events are numbered at the beginning of the interval and the position at impact is indicated by the red circle. Impact of event three occurred before the time series started. The projections of the position on the x-y, x-z and y-z planes are shown in grey.}
 \label{fig:cirgeom}
 \end{figure}

During this interval Rosetta was in bound orbit around the comet at a distance mainly between 10 km and 30 km from the nucleus centre of mass, and mainly in the terminator plane. The trajectory of Rosetta around the times of impacts is shown in Figure \ref{fig:cirgeom} in the cometocentric solar orbital (CSO) reference frame. In the CSO frame the x-axis is directed toward the sun, the z-axis is parallel to the comet's orbital angular momentum vector and the y-axis completes the right-handed system. The intervals covered correspond to the time series to be shown in the following four figures (Figures \ref{fig:eventa}-\ref{fig:eventd}). The times of impact are indicated by red circles. Next we will present RPC measurements from the cometary plasma environment during the impacts of these CIRs. 

\subsection{Event 1: 22 Oct 2014}
\begin{figure}
  \noindent\includegraphics[width=14cm]{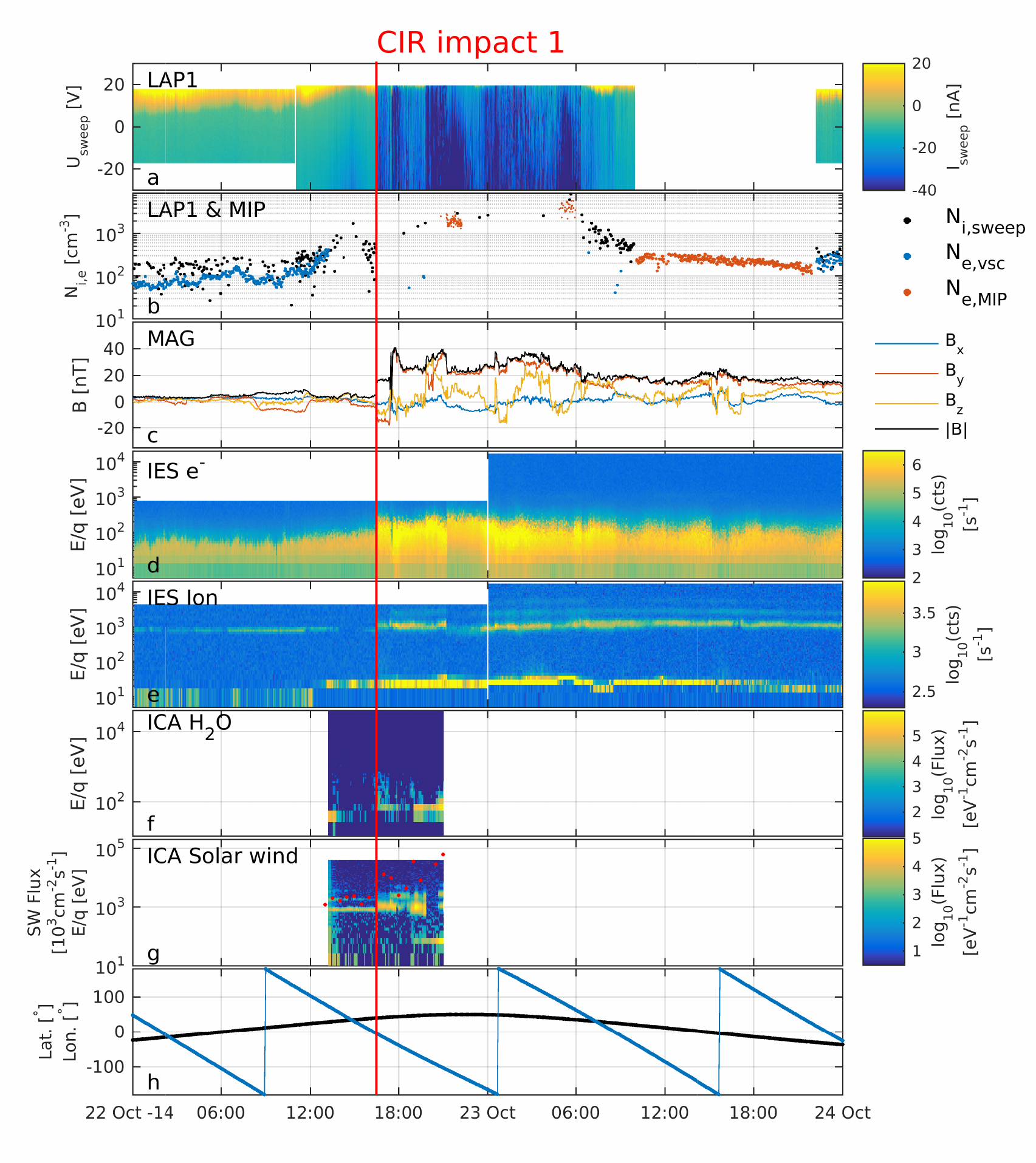}
 \caption{Times series of combined RPC data from the time around the first CIR impact. The panels show (a) LAP bias voltage sweeps (depending on telemetry available the sweep range varies), (b) plasma density estimates from LAP and MIP, (c) vector magnetic field and magnitude from MAG, (d) electron spectrogram from IES, (e) ion spectrogram, (f) spectrogram of cometary water ions from ICA, (g) spectrogram of solar wind ions from ICA  together with solar wind flux (red dots) and finally, (h) cometary longitude (blue line) and latitude (black) of Rosetta. All spectrograms are summed over all elevation and azimuth sectors. The time of impact is indicated by the red vertical line. Note the significant increase in plasma density, magnetic field strength, energetic electron flux and water ion flux at impact.}
 \label{fig:eventa}
 \end{figure}

\begin{figure}
 \noindent\includegraphics[width=14cm]{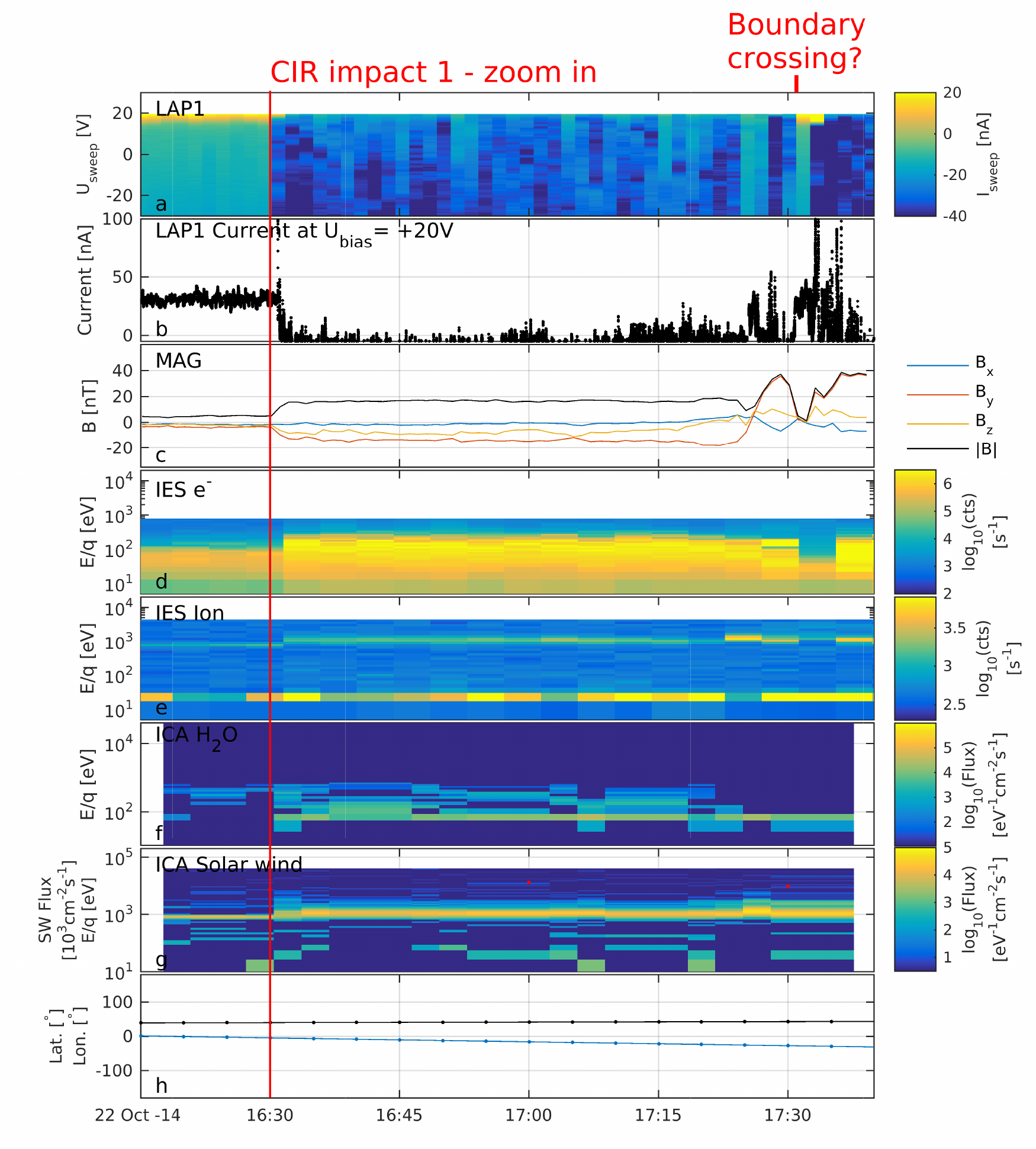}
 \caption{Zoom in from Figure \ref{fig:eventa} during one hour around impact of the first CIR. The format is the same except for that in panel 2 we now show the high-resolution fixed-bias current sampled by LAP1 rather than the density. The bias voltage is +20 V.}
 \label{fig:eventazoom}
 \end{figure}

\begin{figure}
\noindent\includegraphics[width=14cm]{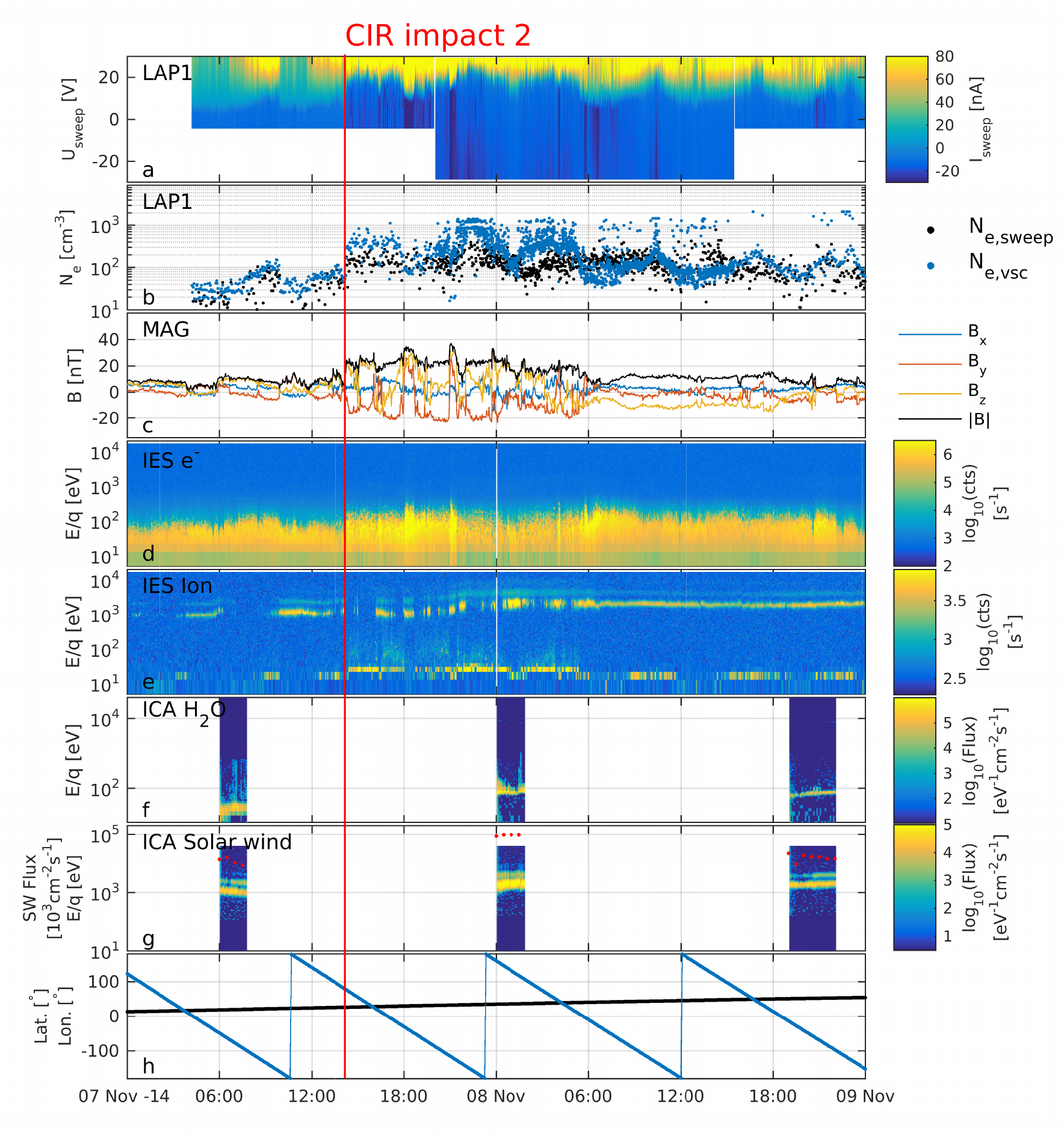}
 \caption{Same as Figure \ref{fig:eventa} but for the second CIR impact. Note again the significant increase in plasma density, magnetic field strength and energetic electrons after impact. The large scale oscillations (time scale of hours) in magnetic field correlates with LAP sweep variations and an increase in the energetic ($\sim$100 eV) electrons from IES.}
 \label{fig:eventb}
 \end{figure}

Figure \ref{fig:eventa} shows a time series of combined RPC data around the time of impact of the first CIR to be studied in detail. Rosetta was at this time in orbit at 10 km from the comet nucleus centre of mass and initially moving to northern latitudes. On 22 Oct 2014 at 16:30 UT, indicated by the red vertical line, a significant disturbance appeared in the plasma environment of comet 67P. In the RPC measurements essentially all parameters shown in Figure \ref{fig:eventa}a-f show significant changes: the LAP ion current (negative voltage side of the sweeps in panel a) suddenly increases by about -20 nA, the MAG magnetic field strength increased from about 5 nT to on average 30 nT in combination with a change of the magnetic field direction, the IES suprathermal ($\sim$10-100 eV) electron counts increased by roughly one order of magnitude together with a general increase in energy of the electrons. The IES-measured solar wind ion energy increased, and the ICA solar wind fluxes (red dots in Figure \ref{fig:eventa}g) increased gradually by almost two orders of magnitude. The count rate of accelerated water ions also suddenly increased. This is identified as the impact of the CIR.

Immediately after impact the spacecraft potential became more negative than the LAP instrument sweep range. The LAP sweep probe potential did not go to sufficiently high positive bias voltage values to be able to attract the electrons through the altered potential field of the spacecraft. Consequently, the electrons could not be measured by the LAP instrument when in sweep mode during about 15 h following impact. The extreme negative spacecraft potential is a clear sign of increased fluxes of electrons to the spacecraft. These increased fluxes of \textcolor{black}{both thermal and suprathermal electrons} significantly disturb the LAP measurements for parts of the interval shown in Fig \ref{fig:eventa}. The strongly negative ion current (panel a, negative voltage), which occasionally goes down below -60 nA, and a non-monotonically increasing ion current during individual sweeps, indicates strong dynamics in the plasma and some might be related to secondary electron emissions caused by the impacting energetic electrons. The LAP ion density can therefore not be estimated for large parts of this interval. 
 
In Figure \ref{fig:eventa}b, we show the ion and electron densities that still can be estimated, from four independent measurements. LAP provides the ion density from the negative-voltage side of the sweeps (when the spacecraft potential is known) as well as the electron density determined from the spacecraft potential measured by LAP. The ion density might be overestimated when the sweep range only goes down to -18V. As the spacecraft potential is driven significantly negative by the energetic electrons, \textcolor{black}{the electron density derived from the spacecraft potential tends to be overestimated}. We therefore chose to be rather conservative and manually lower this estimate by using an electron temperature of 7.5 eV instead of the measured 5 eV\textcolor{black}{ in the calculation of the density}. MIP provides the electron density when in LDL mode during the long interval on 23 Oct, and the electron density when in SDL mode during the two shorter intervals on 22-23 Oct 2014. 

Altogether, these four density estimates paint a rather coherent picture (although with some unfortunate data gaps) of how the local plasma density increases by more than an order of magnitude, from $\sim50-300$ cm$^{-3}$ in the interval before impact to $\sim2000-5000$ cm$^{-3}$ during the interval when the comet is impacted by the CIR. The MIP density agrees rather well with the available LAP ion density estimates in the high-density interval after impact when the spacecraft potential signal is briefly recovered, as well as before and after impact. However, on 23 Oct the MIP density represents a lower bound since the instrument cut-off at 350 cm$^{-3}$ in LDL mode sets in.

The plasma density increases before the identified impact time, already at about 12:00. This is caused by the fact that Rosetta moves toward \textcolor{black}{northern summer latitudes and to over the comet neck region where the neutral outgassing and the plasma density are higher \citep{hassig2015, edberg2015}. The two short intervals, half and one comet rotation later (6h and 12h), when the density is sufficiently high for MIP to provide a density estimate when in its SDL mode (brown dots), also occur above the neck region. Now the density reaches a maximum of 5000 cm$^{-3}$. The neutral gas density does not increase by more than a factor of 5 when moving toward the northern latitudes, while the plasma density increases by at least a factor of 10. So the CIR seems to cause the plasma density to at least double during this interval}. During the previous and the following orbits when at high latitudes again the plasma density does not increase as much as during the CIR impact event.

We also note that both the magnetic field direction (mainly the B$_z$ component) and the energy and flux of suprathermal electrons are observed to vary with a time scale close to half the comet rotation period, indicating that although the solar wind disturbance is large, \textcolor{black}{the interaction region between the cometary plasma environment and the solar wind is still dependent on the structured ionosphere of the comet and its rotation phase}.

After the CIR impact, the cold plasma density and field strength as well as the energetic ion and electron count rates remain high for about a day before they all go back to more normal values, when the CIR has passed. 

In Figure \ref{fig:eventazoom} we show a zoomed in part from Figure \ref{fig:eventa} just at the time of impact of the CIR. Here the density data in panel b has been exchanged with the current measurements from LAP1, sampled at a fixed bias potential of +20 V and with a time resolution of 35 ms. This current actually anti-correlates with plasma density in this case, as higher density as well as increased flux of suprathermal electrons both serve to drive the S/C potential negative, and this effect overcomes the proportionality of current to density. The drop in current after the CIR impact at 16:30 thus means that the LAP probe no longer sees the electrons, only the ions, as the probe (despite a +20 V bias potential) go negative with respect to the plasma. Nevertheless, the fast variations seen after 17:20 indicate corresponding rapid variations in the plasma, at time scales not possible to resolve in the particle data. An interesting feature occurs at 17:31-17:33 UT, when, simultaneously, the electron current in panel b increases (meaning that the density decreases) for 2 minutes, the magnetic field strength decreases (while still fluctuating as seen in high resolution MAG data - not shown), the energetic electron count decreases significantly, to values lower than before the CIR impact, and the solar wind ion counts decrease significantly. Also the electron current in the LAP sweeps is momentarily retrieved as the S/C potential decreases. These could possibly be the signatures of a plasma boundary forming as the CIR impacts, \textcolor{black}{which Rosetta moves across}. We will discuss this possibility further after having presented the remaining three CIR events.
 
\subsection{Event 2: 7 Nov 2014}
Figure \ref{fig:eventb} shows a time series of the same format as \textcolor{black}{Figure \ref{fig:eventa}} but covering the interval of the second CIR impact. The second CIR is observed to impact at 14:15 UT on 7 Nov 2014 and again causes a significant disturbance to the plasma environment. Rosetta was at this time in orbit at 30 km from the centre of mass, and in the northern \textcolor{black}{illuminated (summer) and more active} hemisphere. Both the neutral gas density and the plasma density are naturally lower at this time than when in orbit at 10 km. The CIR impact signatures observed by RPC are similar to those during the previous event: the magnetic field strength and energetic electron fluxes increase suddenly, the LAP ion current changes significantly by about -10 nA, and the spacecraft potential becomes more negative. The cold plasma density more than doubles, from $\sim50$ cm$^{-3}$ before impact to about $\sim300$ cm$^{-3}$ immediately after impact, following the sweep-derived electron density. The density estimated from the spacecraft potential probably overestimates the value due to increased fluxes of suprathermal electrons, even though we assume an electron temperature of 10 eV during this event. The electron density from the sweep might at the same time be underestimated due to the negative spacecraft potential. MIP did not measure any electron plasma oscillations in this interval when it was run in SDL mode, which suggests that the Debye length was much larger than a few tenth of cm. This is consistent with a plasma density that stays below 1000 cm$^{-3}$ for 5-10 eV electrons. Some of the LAP observed density increase is attributed to going toward \textcolor{black}{higher and more illuminated latitudes}, but since the latitude continues to increase after the CIR has passed, and the density goes back to nominal values, it is clear that some of the density increase \textcolor{black}{(at least a factor of 2 if being conservative)} is caused by the CIR impact. 

The magnetic field strength during this event increases from about 10 nT to 30 nT around the time of impact. The IES suprathermal electron fluxes and ion fluxes increases after impact, similar to the previous event. There are also low-energy ions (10-100 eV) appearing after impact, which might only come into view of the instrument as the magnetic field direction (and consequently the electric field) changes. ICA was on during three short intervals of this event. However, during the passing of the CIR, ICA did measure the solar wind fluxes to be one order of magnitude higher than during the two short intervals prior to and after the CIR impact. 

There are also significant variations in the magnetic field orientation in this interval, which are simultaneous with the bursts of $\sim$100 eV electrons measured by IES (bright yellow patches centered at 16:30, 18:20 and 21:20 UT on 7 Nov 2014 and at 06:00 on 8 Nov 2014).  The ion current from the LAP probe also increases at these instances. The increased suprathermal electron fluxes during certain magnetic field orientations could indicate that the electrons are heated through the solar wind interaction and accelerated in the direction of the magnetic field. It could also be that the properties of the CIR itself are variable with bursty dynamic pressures enhancements and with a changing IMF direction. The intermittent solar wind signal in IES ion data is partially correlated with the electron and magnetic field signatures as well, which is probably due to the solar wind deflection changing as the IMF changes direction \citep{broiles2015}. 

 \subsection{Event 3: 3 Dec 2014}
\begin{figure}
\noindent\includegraphics[width=14cm]{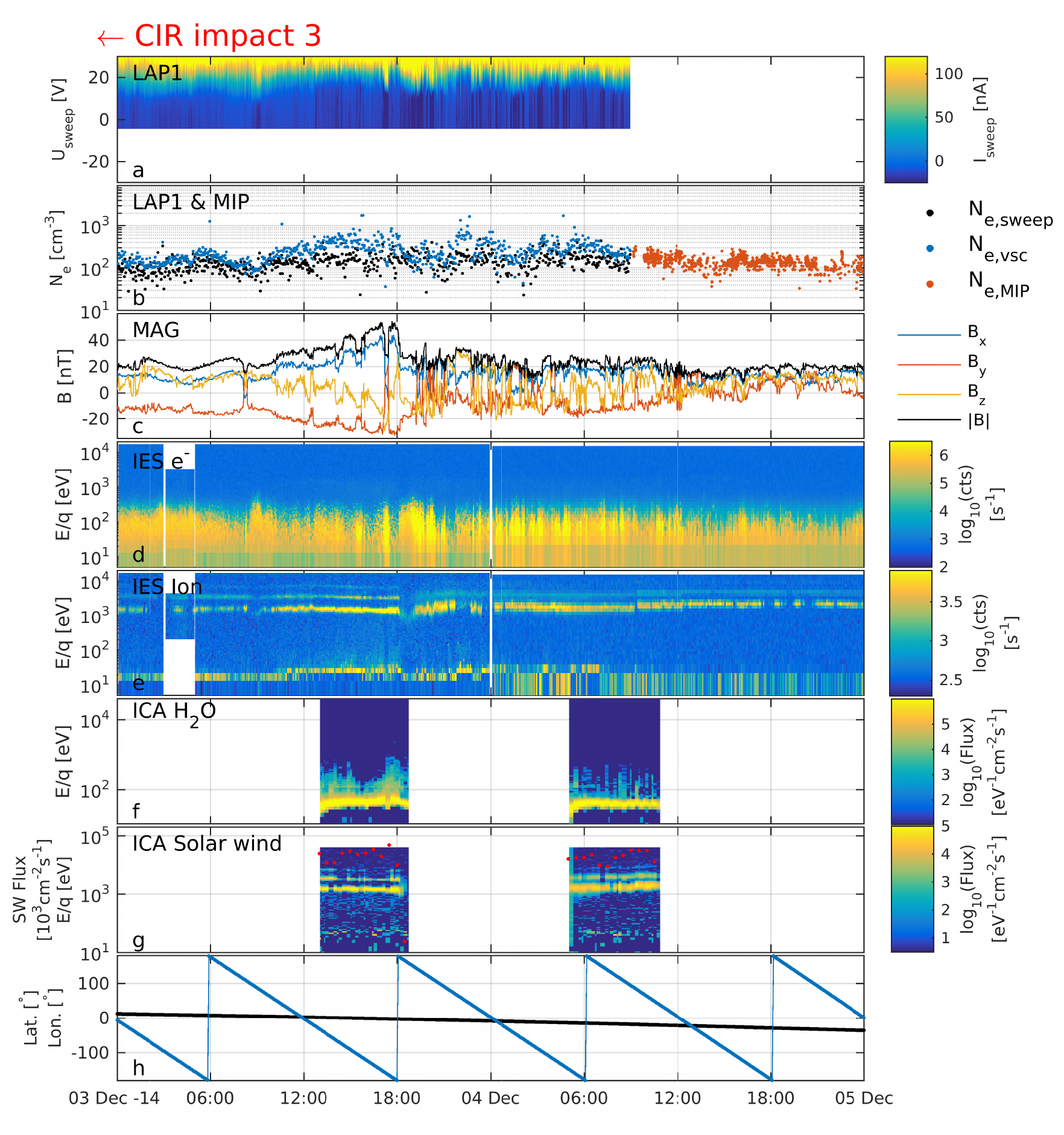}
 \caption{Same as Figure \ref{fig:eventa} but for the third CIR event. The initial CIR impact occurs already on 27 Nov 2014, but continues over several days and covers the interval shown here. In the interval shown here (still during the passing of the CIR) there is a gradual increase in both plasma density and magnetic field strength, which is followed by a sudden drop in magnetic field strength and an increase in magnetic field fluctuations over the following 24 h. At the same time the 100 eV electron flux is further increased compared to quiet solar wind times.}
 \label{fig:eventc}
 \end{figure}

The third event is shown in Figure \ref{fig:eventc} in the same format as before. Rosetta was at this time again at a distance of 30 km from the comet nucleus but moving slowly toward southern latitudes. This time the impact of the CIR is much more gradual. There is a less sharp shock front impacting on the comet ionosphere and already on 27 Nov 2014 the solar wind started to increase, as can be seen in the longer overview plot in Figure \ref{fig:sw}e. However, we will focus on the interval starting on the 3 Dec 2014, when a second pulse of high velocity solar wind reaches the comet. The plasma density increases gradually after the initial impact on 27 Nov, from $\sim50$ cm$^{-3}$ to values around 200 cm$^{-3}$. The density estimate from the spacecraft potential may overestimate the values again in this interval even though we have raised the electron temperature in the density estimate model to 10 eV. The general density increase occurs when Rosetta is moving toward northern latitudes, where the density should be higher, but on the other hand, the density was about a factor of 2 lower two weeks earlier when previously in the northern hemisphere, meaning that the CIR impact causes a factor of 2 increase in the plasma density. The MIP and LAP density estimates match quite well during this interval.

Similar to the previous event, the IES measured fluxes of energetic electrons (yellow bright patches in Figure \ref{fig:eventc}d) appear as very bursty and are correlated with the magnetic field oscillations, and also with the LAP ion current increases (interpreted as a combination of increased density and secondary electron emission). 

A unique feature of this event is the magnetic field signature. At the same time as the density gradually increases, the magnetic field strength is also gradually increasing, from 20 nT at the start of the interval to 50 nT at 18:00 UT on 3 Dec 2014. After this the field strength sharply decreases back to 25 nT. At the same time the solar wind ions disappear from the IES and ICA ion measurements (Figure \ref{fig:eventc}e and g) and are not seen from 18:10 - 19:10 UT, likely due to deflection of the solar wind \citep{broiles2015}. The ICA solar wind flux measurements (red dots in Figure \ref{fig:eventc}g) provide one data point close to when the magnetic field strength drops to 25 nT, which shows a value two orders of magnitude lower than before, consistent with the solar wind disappearing. The magnetic field strength drop occurs while Rosetta is moving closer to the comet and the distance decreases steadily from 30 km to  25 km over the day (see Figure \ref{fig:sw}f). After the drop in magnetic field strength there are rapid magnetic field fluctuations, which have not been observed during any of the other events, and the plasma density decreases somewhat. These signatures could be interpreted as the crossing of a plasma boundary, which appears as the solar wind increase compresses the cometary plasma environment. During the 6 hours prior to this major decrease in magnetic field strength, there are four short ($\sim10$ min) drops in the magnetic field strength, which could then be interpreted as the boundary moving back and forth as the solar wind pressure varies. \textcolor{black}{It could also be that these are signatures of strong dynamics in the plasma environment, a possibility we will expand on further in the Discussion section}.

 \subsection{Event 4: 22 Dec 2014}
\begin{figure}
\noindent\includegraphics[width=14cm]{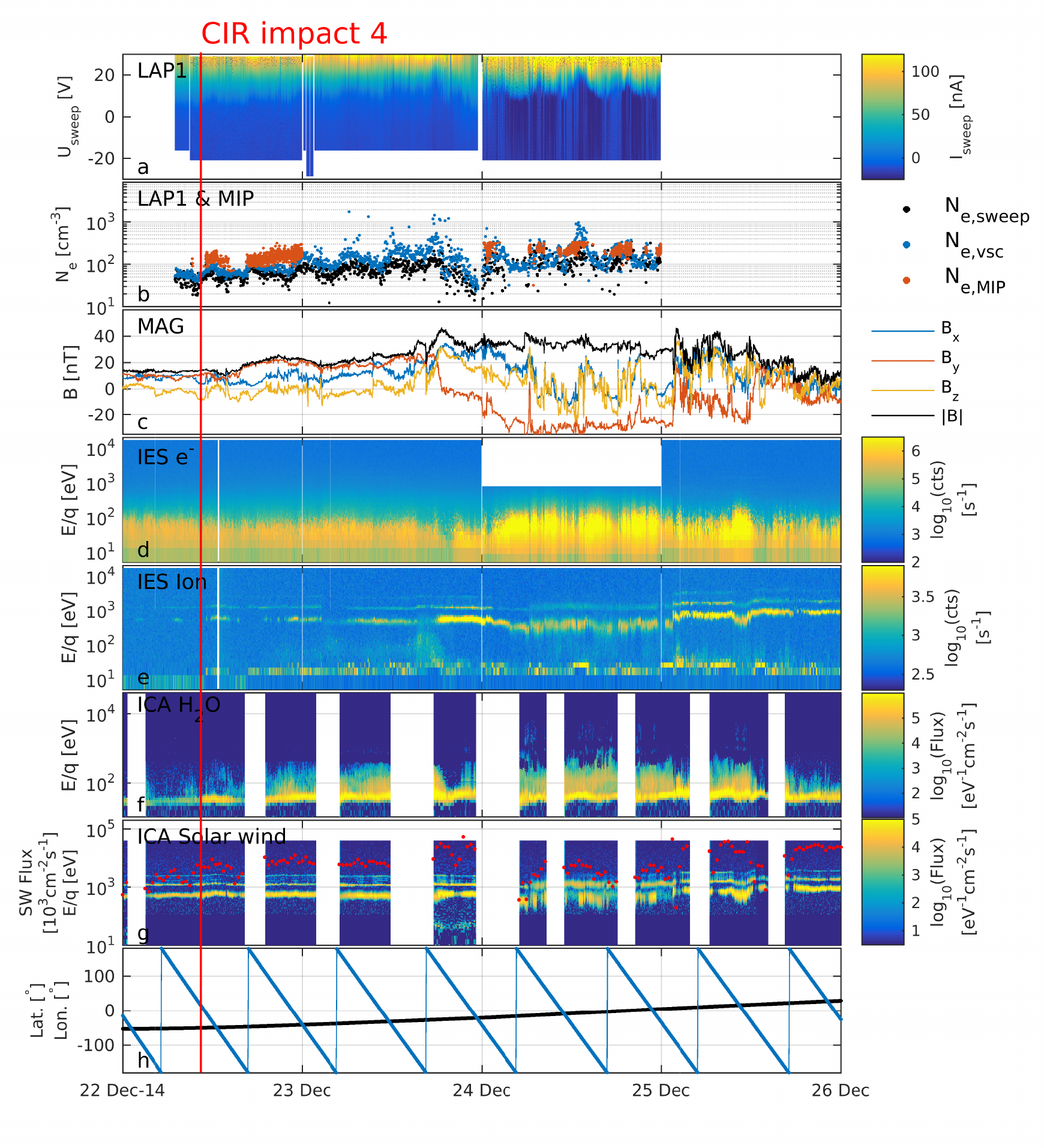}
 \caption{Same as Figure \ref{fig:eventa} but for the fourth CIR impact. Note the large scale change in the magnetic field orientation and the modest increase in plasma density. \textcolor{black}{There is an increase in density immediately after the CIR impact, particularly seen as that is when MIP starts observing features at the plasma frequency in the mutual impedance spectra to provide density estimates. This is consistent with a discontinuity toward a smaller Debye length at the CIR impact}. The energetic electrons still appear in this interval but are most prominent 1.5 days after impact, at the same time as the solar wind flux decreases. Water ion fluxes of energies up to 100 eV are elevated during the entire interval.}
 \label{fig:eventd}
 \end{figure}

The fourth and final CIR impact also occurs in two steps similar to the previous event, and is shown in Figure \ref{fig:eventd}. The initial impact happened on 22 Dec 2014 at 09:00 UT when Rosetta again was at a distance of 30 km from the comet centre of mass. At impact, the plasma density shows a moderate increase, which is simultaneous with moving towards higher latitudes. The density still increases to a higher value than during the previous pass over the same latitude region, \textcolor{black}{indicating that the CIR again causes the density to increase}. The magnetic field strength increases gradually, from about 20 nT to 45 nT, similar to the previous event. The increase lasts until about 20:00 UT on 23 Dec 2014, after which the field strength slowly decreases and the magnetic field orientation slowly changes. On 24 Dec 2015, when the magnetic field orientation has changed, the solar wind flux is decreased and the suprathermal electrons increased significantly. It is possible that these signatures also indicate the crossing of a plasma boundary, which builds up as the solar wind dynamic pressure increases. At 02:00 UT on 25 Dec 2014, the field strength increases briefly and the field changes orientation suddenly. 

ICA was fortunately on for most of the time during this event and measured increased fluxes of accelerated water ions from the time of impact until noon on 25 Dec 2015, when the CIR had passed (Figure \ref{fig:eventd}f). This also appeared to be the case during CIR event 1, when ICA was on during a few hours around impact. The accelerated water ion fluxes are increased during the entire passing of the CIR and only decreases toward the end of 25 Dec 2015. The high fluxes are observed even though the magnetic field changes direction. These accelerated ions are interpreted as the population of pick up ions by the solar wind induced $\textbf{V} \times \textbf{B}$ field. As gradient lengths are small compared to the ion gyro-radius, almost all ions will only have seen part of this E-field, giving a broad energy distribution \citep{nilsson2015b}. 

During the second impact of this event, on 27 Dec 2014, LAP was unfortunately not switched on and MIP was operated in SDL mode, which is blind to 5-10 eV electrons of few hundred particles per cm$^{-3}$. Still, the signatures in the data from the rest of the RPC instruments (not shown) suggest a similar behaviour as during the other events, except that the magnetic field strength as well as fluxes of energetic electrons are higher in the 27 Dec event, and the energetic electrons reach energies of several hundreds of eV. 

\section{Discussion}
During the CIR impact events presented here the cold plasma density is seen to increase \textcolor{black}{significantly}, but the response is different \textcolor{black}{for each} event. During the first event, the cometary ionospheric density is clearly seen to increase, to above 1000 cm$^{-3}$ at 10 km when the CIR impacts, and during the second event the density increases to $\sim$300 cm$^{-3}$ at 30 km. During the third and fourth events, the density increase is more modest but still significant. Hence, each event seems to influence the cometary plasma environment slightly different, in terms of increasing the cold plasma density. 

The cause of the significant enhancements of the low-energy plasma could be compression of the local plasma by the increased solar wind, particle impact ionisation and/or charge exchange processes when the increased flux of solar wind plasma impacts on the coma. It could also be that more mass-loaded solar wind is accelerated in the direction of the comet ionosphere. The relative increase in the magnetic field strength during event 1 is $\sim$5, while the relative increase in the density is $\sim$ 4-7. For event 2 the relative increases are $\sim$4 and $\sim$2.5, respectively. The similar relative increase between the magnetic field strength and the density suggests that compression of the plasma is likely occurring. \textcolor{black}{Furthermore, as the compression leads to an increased density of both the thermal and suprathermal plasma, the ionization rate through particle impact and charge exchange also goes up. An increased ionisation rate gives higher ion pick up rate, which leads to a slowing down of the solar wind. The plasma then gets further compressed as the trailing solar wind catches up, which in turn leads to even more increase in both the density and the magnetic field strength, i.e. the same signatures as seen during direct compression by an increased solar wind dynamic pressure}. 

\textcolor{black}{Comparison between ionospheric models including compression of the plasma and measurements of the suprathermal electrons on 23 Oct (during CIR event 1) by Madanian et al., (submitted) show good agreement, which then supports that explanation. ICA does not confirm increased charge exchange (no major increase in He$^+$ fluxes is observed at this time). The increase in suprathermal electron fluxes ought to cause some increased particle impact ionisation, but to what extent and how much that would increase the density would require further modelling, which is beyond the scope of this paper.}

The fact that we seem to see a less dramatic increase in the low-energy plasma density during the later events compared to the first event could be because the outgassing increases and the cometary coma becomes more dense when approaching the Sun. A denser coma \textcolor{black}{makes compression harder, and} hinders more incoming flux from upstream, such that less solar wind plasma reaches the deep coma where Rosetta was located. As the coma grows, deflection of solar wind plasma also increases due to stronger electric fields \citep{broiles2015}, resulting in less solar wind plasma reaching the near-nucleus environment \citep{nilsson2015b}. Also, the solar wind increase is less dramatic during the initial impact in event 4 compared to events 1-3, and we lack LAP data during the main impact.

We stress that the density estimates presented here are not without uncertainties and there are several possible sources of errors. These include an uncertain photoelectron current estimate, additional secondary electron emission or a hotter electron temperature and finally, the LAP sweep range being limited and missing parts of the plasma population. Still, using different density estimates from both LAP and MIP, which present rather similar values, provides confidence to the density estimates, and the uncertainty of the relative increase of the density in response to the CIR impact and between the events is nevertheless still quite satisfying.

For the first CIR event, a related observation was presented by \citet{feldman2015}. They showed measurements from the ALICE far-ultraviolet spectrograph of an increased brightness in the comet limb spectrum on the 22 Oct 2014, i.e. at the time of impact of the first CIR. This was attributed to an increased level of photoelectron impact dissociation of CO$_2$.

During all four CIR impact events, the fluxes of $\sim$10-100 eV electrons are seen to increase significantly. The presence of these electrons, and especially the cause of their energy, is so far unclear \citep{clark2015}. They must be accelerated and heated through some mechanism since 100 eV is much higher than what is expected after, for instance, photoionisation (10-15 eV) of neutral species. Here we can report that the acceleration of this population is clearly related to the solar wind interaction since the suprathermal electrons increase in both energy and flux during intervals of increased solar wind energy input. Furthermore, the measured variations in the electron fluxes are apparently connected to the magnetic field orientation. These energetic electrons also impact on the spacecraft and drive the spacecraft potential negative, and possibly distort the plasma density measurements by LAP. 

Another effect that the CIRs might have is that the sputtering of the comet nucleus \citep{wurz2015}, as well as of larger dust grains, increases as the solar wind particle flux increases. However, this probably does not significantly increase the total plasma density.

During event \textcolor{black}{1, 3, and 4 we observe some unusual signatures in the plasma environment, which are challenging to explain decisively from single spacecraft measurements. In the following paragraphs we will describe these more carefully and discuss if the signatures are either those of plasma boundaries forming in the comet environment, or perhaps the ionospheric response to tail disconnection events.} During the first event (Figures \ref{fig:eventa} and \ref{fig:eventazoom}), about an hour after impact, the magnetic field strength increases, the spacecraft potential becomes less negative so that the measured plasma density decreases and the LAP continuous electron current increases, while the suprathermal electron flux decreases to values lower than before the impact. This could be the signature of skimming \textcolor{black}{a plasma boundary (such as an ionopause or contact surface, for instance), which could be formed briefly at 10 km from the nucleus during a period when the upstream solar wind dynamic pressure is higher than usual}. Alternatively, this could be a temporary decrease in the solar wind dynamic pressure, which simply relaxes the disturbance of the cometary plasma environment. However, the fact that both the suprathermal electron fluxes and the magnetic field decreases to values lower than before the impact indicates against this explanation. Also, the LAP sweep during the minute after stands out from all other sweeps during the day, as the spacecraft potential gets less negative than previously, which is consistent with lower density and lower flux of energetic electrons.

During event 3 (Figure \ref{fig:eventc}) the magnetic field gradually piles up during several hours after the CIR impact. Then the field strength drops sharply at 18:00 on 3 Dec 2014. Immediately after this drop the low-energy plasma density decreases at the same time as a burst of energetic electrons is observed. This could possibly be \textcolor{black}{the signatures of the ionospheric response} of a tail-disconnection event, when magnetic flux is piled-up upstream of the comet before being released during periods of increased solar wind flow. The energetic electrons could have been accelerated \textcolor{black}{into the ionosphere} as the piled up magnetic flux is released. \textcolor{black}{A tail-disconnection event could also occur when the diamagnetic cavity is formed, to pre-condition the tail with strong anti-parallel magnetic fields. Although no diamagnetic cavity has been observed in this interval, it is possible that it forms briefly close to the comet following a solar wind pressure pulse.} 

Another possibility is that this is also a plasma boundary forming and as Rosetta moves closer to the comet nucleus and the solar wind increases, the boundary is crossed and Rosetta is entering another plasma region. During event 4, similar signatures are also observed in both the magnetic field and plasma data as the field gradually piles up and a region characterised of high suprathermal fluxes, lower solar wind fluxes and increased fluxes of cometary accelerated water ions is entered. Alternatively, this could again simply be a relaxation of the impinging solar wind dynamic pressure, although then the transition is perhaps sharper than one would expect from a relaxing solar wind. \textcolor{black}{Further work, including modelling and studies of additional events, are required before these observations can be firmly explained.}

We have shown RPC data from four CIR impacts on comet 67P, during three months in 2014, when the comet was relatively inactive. After these, more events have been observed in RPC data on the dates 14 Jan, 29 Jan, 16 Feb, 24 Feb, 6 Mar, 13 Mar, 17 Mar and 21 Mar 2015. These events are either CIRs of similar kind as shown in the four examples here or CMEs. In April 2015 the solar wind signal was finally lost in the RPC data since at this time it was being completely shielded/deflected off by the growing cometary coma. Since the cometary activity is constantly increasing the signatures of these events will gradually change and we chose not to study them in any detail here, but rather leave them for a future separate study.

\section{Summary}
CIRs cause a significant disturbance to the cometary plasma environment upon impact. The properties of individual CIRs vary and the response of the cometary plasma environment consequently differs from event to event. We have studied the impact of four CIRs on comet 67P, after having traced them from Earth via Mars and finally to 67P using solar wind monitoring measurements. These solar wind pressure pulses are observed in the interval 22 Oct 2014 to 25 Dec 2014, when the comet was at a heliocentric distance of 3.1 to 2.7 AU. Rosetta was at this time orbiting the comet at a distance of 10-30 km from the nucleus centre of mass. The enhanced solar wind dynamic pressure together with changes in IMF strength and orientation, are the cause of a number of features in the cometary plasma environment out of which the most prominent are outlined below:

- the cold (few eV) ionospheric plasma density increases by a factor of at least 2 during three events, due to compression of the plasma environment, increased particle impact ionisation and possibly charge exchange processes\textcolor{black}{, which in turn leads to an increased ion pick up rate}.

- energetic (10-100 eV) electrons, which appears to be a ubiquitous population around 67, are significantly increased in both flux and energy, indicating that these electrons are heated through the solar wind interaction with the cometary plasma. The fluxes vary with changes in the magnetic field direction.

- the spacecraft potential drops to values typically below -20 V\textcolor{black}{, as the flux of electrons increase}.

- accelerated cometary water ions increase in both energy and flux.

- the magnetic field piles up around the comet and increases by a factor of 2-5 to reaches values around 50 nT at maximum.

- the increased dynamic pressure possibly causes a short-lived (2-3 min) plasma boundary to appear on 22 Oct 2014, characterised by a drop in the magnetic field strength, drop-out of energetic electrons and an unusually steady electron current to the LAP probes, followed by an increased electron density the minute after.

- on the event on the 3 Nov 2014 the piled up magnetic flux is suddenly released as the magnetic field strength sharply decreases. This is followed by significant magnetic field fluctuations and increased fluxes of energetic electrons over the following 24h. These might be the signatures of a magnetic pile-up boundary having formed \textcolor{black}{or possibly the ionospheric response to a tail disconnection event}. Similar features are observed during the fourth CIR event, on 23-25 Dec 2015.

In conclusion, the impacting CIRs produce a very dynamic and variable interaction with the rotating comet and its spatially structured and time-varying neutral gas outflow. The magnetic field orientation and strength, energetic electron fluxes, accelerated cometary water ion fluxes and cold plasma density are all varying extensively in the cometary plasma environment. Depending on these properties the CIR interaction with the comet ionosphere will vary accordingly.


%
%
%
%
%
%
%

\begin{acknowledgments}
Rosetta is a European Space Agency (ESA) mission with contributions from its member states and the National Aeronautics and Space Administration (NASA). The work on RPC-LAP data was funded by the Swedish National Space Board under contracts 109/02, 135/13, 166/14 and 114/13 and Vetenskapsr\aa det under contracts 621-2013-4191 and 621-2014-5526. This work has made use of the AMDA and RPC Quicklook database to provide an initial overview of the events studied. This is provided through a collaboration between the Centre de Donn\'ees de la Physique des Plasmas (CDPP) (supported by CNRS, CNES, Observatoire de Paris and Universit\'e Paul Sabatier, Toulouse) and Imperial College London (supported by the UK Science and Technology Facilities Council). \textcolor{black}{We thank K.C. Hansen and B. Zieger for providing solar wind propagations from their Michigan Solar Wind Model (http://mswim.engin.umich.edu/).} The data used in this paper will soon be made available on the ESA Planetary Science Archive and is available upon request until that time.
\end{acknowledgments}

\end{article}
%
%
%
%
%
%
 
 \clearpage
 
%
%


\end{document}